\documentclass[12pt]{article}

\input amssym.tex

\begin{document}

\title{The quantum theory of the free Maxwell field on the de Sitter
expanding universe}

\author{Ion I. Cot\u aescu \thanks{E-mail:~~cota@physics.uvt.ro}\,
and Cosmin Crucean \thanks{E-mail:~~crucean@physics.uvt.ro}\\
{\small \it West University of Timi\c soara,}\\
{\small \it V. Parvan Ave. 4 RO-300223 Timi\c soara,  Romania}}

\maketitle

\begin{abstract}
The theory of the free Maxwell field in two moving frames on the de
Sitter spacetime is investigated pointing out that the conserved
momentum and energy operators do not commute to each other. This
leads us to consider new plane waves solutions of the Maxwell
equation which are eigenfunctions of the energy operator. Such
particular solutions complete the theory in which only the solutions
of given momentum were considered so far. The energy eigenfunctions
can be obtained thanks to our new time-evolution picture proposed
previously for the scalar and Dirac fields. Considering both these
types of modes, it is shown that the second quantization of the free
electromagnetic potential in the Coulomb gauge can be done in a
canonical manner as in special relativity. The principal conserved
one-particle operators associated to Killing vectors are derived,
concentrating on the energy, momentum and total angular momentum
operators.

Pacs: 04.62.+v
\end{abstract}


\newpage

\section{Introduction}

The quantum theory of fields on curved spacetimes deals with quantum
systems in the presence of gravitation but without to affect the
background geometry. Of a special interest in cosmology is the de
Sitter (dS) expanding universe carrying fields variously coupled to
gravity. The free fields (minimally coupled) are the principal
ingredients in calculating scattering amplitudes using
perturbations.

The large symmetry of the dS geometry provides us with then Killing
vectors giving rise to corresponding isometry generators. These are
conserved operators in the sense that they commute with the
operators of the field equations. Our theory of external symmetry
\cite{rf:1} enables one to write down the isometry generators for
fields of any spin applying a generalized Carter and McLenaghan
formula \cite{rf:1,rf:2}. This holds in local frames but even in
natural ones (i. e. local charts) as we have shown recently
\cite{rf:3}. The presence of the isometry generators offer us the
opportunity to select suitable sets of commuting operators able to
determine the quantum modes as common eigenfunctions.

There exists one Killing vector which is time-like but only in the
observer's light-cone. Despite of some doubts appeared in literature
\cite{rf:4} this  vector was used by many authors for defining the
energy (or Hamiltonian) operator of the dS spacetime \cite{rf:5,rf:6}.
This operator commutes with the components of the total angular
momentum being helpful for deriving vector \cite{rf:5} and Dirac
\cite{rf:6} quantum modes in the dS static chart with spherical
coordinates. However, in the moving charts with proper or conformal
times, the same energy operator depends on the space coordinates and
their derivatives so that this does not commute with the momentum
components \cite{rf:7}. Consequently, the energy and momentum
operators can not be put simultaneously in diagonal form. This means
that these operators must be included in two different sets of
commuting operators determining two different bases as sets of
common eigenfunctions. The basis in which the momentum is diagonal,
called here the {\em momentum basis}, is well-studied for various
fields \cite{rf:8,rf:9} of course. However, the basis in which the
energy is diagonal is interesting because it relies on  new
particular solutions of the field equations which should be
eigenfunctions of the energy operator.

In order to derive these solutions we proposed a new time-evolution
picture in moving dS charts, called the Schr\" odinger picture,
where new quantum modes of determined energy and momentum direction
can be derived \cite{rf:10} for the scalar \cite{rf:11} and  Dirac
\cite{rf:12} free fields. So we obtained the {\em energy bases} of
these fields which are equivalent to the corresponding momentum ones
through  unitary transformations. Both these types of bases have
suitable orthogonality and completeness properties allowing us to
perform the canonical quantization of the scalar \cite{rf:11}, Dirac
\cite{rf:7} and (massive) Proca \cite{rf:13} free fields on dS
manifolds. Moreover, we pointed out that the momentum and energy
bases yield the {\em same} vacuum state which is analogous to the
Bunch-Davies vacuum of the scalar field \cite{rf:14}. In this
framework we obtained new Dirac \cite{rf:7} and Proca \cite{rf:13}
propagators but by recovering the well-known one of the scalar field
\cite{rf:15}. However, the principal virtue of our approach is the
opportunity of calculating the conserved one-particle operators
corresponding to the Killing vectors of the dS geometry now in both
the momentum and energy bases.

In the present paper we would like to study the free Maxwell field
on the dS spacetime in a similar approach. Our main purpose is to
use the canonical quantization in order to establish  the form of
the principal one-particle operators. The central point is to derive
the mode functions of the momentum and energy bases in the moving dS
charts with proper or conformal times \cite{rf:8}. Taking into
consideration the conformal invariance of the Maxwell equation, we
observe that the Coulomb gauge is the only gauge which opens the way
to conformally relate the whole theory written in the chart of
conformal time to the usual electrodynamic of special relativity.
Exploiting this conjecture we obtain the mode functions of the
momentum basis in Coulomb gauge as in the flat case \cite{rf:8}.
Furthermore, keeping the same gauge, we use the Schr\" odinger
picture for deriving the new quantum modes of the energy basis
determined by energy, momentum direction and polarization. In this
framework we perform the canonical quantization in Coulomb gauge as
in QED. Hereby we deduce the mode expansions of the principal
one-particle operators, i. e. the energy, momentum, total angular
momentum and polarization. In what concerns the problem of
propagators (or two-point functions \cite{rf:15,rf:16}) we bring nothing
new since the Maxwell propagators in the chart of conformal time
have the same form as in special relativity. We note that the theory
of the Maxwell field in Coulomb gauge presented here is the natural
massless limit of the theory of the Proca field on the dS expanding
universe \cite{rf:13}.

In the second section we begin with a brief review of the theory of
the free Maxwell field on the mentioned moving dS charts,
introducing the principal conserved operators and defining the
Schr\" odinger time-evolution picture. This allows us to find the
new quantum modes of the energy basis as well as the transition
coefficients between these modes and those of the momentum basis.
Section 4 is devoted to the canonical quantization of the free
Maxwell field in Coulomb gauge which provides us with well-known
Green functions but depending now on the conformal time. The mode
expansions of the principal one-particle operators are written down
in section 5.

We must stress that our principal purpose is to construct a rigorous
mathematical theory of the canonical quantization of the free
Maxwell field on the dS expanding universe rather than to explore
the problems of actual interest in cosmology \cite{rf:17,rf:18}.

\section{The Maxwell free field on de Sitter spacetimes}

Let $(M,g)$ be a curved spacetime and $\{x\}$ a (local) chart of coordinates
$x^{\mu}$ ($\mu,\nu,...=0,1,2,3 $) and the line element
\begin{equation}
ds^{2}=g_{\mu \nu}d{x}^{\mu}d{x}^{\nu},
\end{equation}
defined by the metric tensor $g_{\mu \nu }$. We denote by $A$ the
(electromagnetic) potential minimally coupled to gravity, whose action reads
\begin{equation}\label{action}
\mathcal{S}[A]=\int d^{4}x \sqrt{g}\,{\cal L}=-\frac{1}{4}\int d^{4}x
\sqrt{g}\,F_{\mu \nu  }F^{\mu \nu  } ,
\end{equation}
where $g=|\det(g_{\mu\nu})|$ and $F_{\mu \nu  }=\partial_{\mu } A_{\nu
}-\partial_{\nu } A_{\mu }$ is the field strength. From this action one derives
the field equation
\begin{equation}
\partial_{\nu}(\sqrt{g}\,g^{\nu\alpha}g^{\mu\beta}F_{\alpha\beta})=0\,,
\end{equation}
which is invariant under conformal transformations, $g_{\mu\nu}\to
g_{\mu\nu}'=\Omega g_{\mu\nu}$ and $A \to A'$ where
\begin{equation}\label{conf}
A'_{\mu}=A_{\mu}\quad A^{\prime\,\mu}=\Omega^{-1} A^{\mu}\,.
\end{equation}
The canonical variables $A_{\mu}$ must obey, in addition, the
Lorentz condition
\begin{equation}\label{Lor}
\partial_{\mu}(\sqrt{g}\,g^{\mu\nu}A_{\nu})=0\,,
\end{equation}
which is no longer conformally invariant since
\begin{equation}\label{Lor1}
\partial_{\mu}(\sqrt{g'}\,g^{\prime\,\mu\nu}A_{\nu}')=\partial_{\mu}(\sqrt{g}\,g^{\mu\nu}A_{\nu})+
\sqrt{g}A^{\mu}\partial_\mu\Omega\,.
\end{equation}

In general, the whole theory is invariant under symmetry
transformations. Since $A$ is a real field there are no internal
symmetries remaining thus only with the isometries related to the
Killing vectors of $M$. Given an isometry transformation $x\to
x'=\phi_{\xi}(x)$ depending on the group parameter $\xi$ there
exists an associated Killing vector field,
${K}=\partial_{\xi}\phi_{\xi}|_{\xi=0}$ (which satisfy the Killing
equation ${K}_{\mu;\nu}+{K}_{\nu;\mu}=0$). The vector field $A$
transforms under this isometry as $A\to A' =T_{\xi}A$, according to
the operator-valued representation $\xi\to T_{\xi}$ of the isometry
group defined by the well-known rule
\begin{equation}
\frac{\partial \phi^{\nu}_{\xi}(x)}{\partial
x_{\mu}}\left(T_{\xi}A\right)_{\nu}[\phi(x)]=A_{\mu}(x)\,.
\end{equation}
The corresponding generator, $X_{K}=i\,\partial_{\xi}T_{\xi}|_{\xi=0}$, has the
action
\begin{equation}\label{XA}
(X_{K}\, A)_{\mu}=-i({K}^{\nu}A_{\mu;\nu}+{K}^{\nu}_{~;\mu}A_{\nu})\,.
\end{equation}
Notice that this formula can be put in the Carter and McLenaghan
form using  point-dependent spin generators \cite{rf:3}.

We say that the generators $X_K$ are conserved operators since they
commute with the operator of the field equation \cite{rf:1}. Moreover,
according to the Noether theorem, it results that each Killing
vector $K$ gives rise to the time-independent quantity,
\begin{equation}\label{Ck}
C[{K}]=-\frac{i}{2}\int_{\Sigma}d\sigma^{\mu}\sqrt{g}\,g^{\alpha\beta}\left[
A_{\alpha}\stackrel{\leftrightarrow}{\partial_{\mu}}
(X_{K}\,A_{\beta})\right]\,,
\end{equation}
on a given space-like hypersurface $\Sigma \subset M$. We note that this
formula, written with the notation
$f\stackrel{\leftrightarrow}{\partial}g=f(\partial g)-g(\partial f)$, holds
only when Eq. (\ref{Lor}) is fulfilled and the boundary conditions allow one to
apply the Green theorem.

In what follows  $(M,g)$ will be the dS spacetime defined as a
hyperboloid of radius $1/\omega$ \footnote{We denote by $\omega$ the
Hubble constant of the dS spacetime since $H$ is reserved for the
energy operator} in a five-dimensional pseudo-Euclidean manifold,
$M^5$, of coordinates $z^A$ labeled by the indices $A,\,B,...=
0,1,2,3,5$. The local charts of coordinates $\{x\}$ on $M$ can be
easily introduced giving the functions $z^A(x)$. We consider here
either the chart $\{t,{\bf x}\}$ with the {\em proper} time $t$,
Cartesian coordinates and the FRW line element
\begin{equation}\label{mrw}
ds^2= dt^2-e^{2\omega t} (d{\bf x}\cdot d{\bf x})\,,
\end{equation}
or the chart $\{t_{c},\bf{x}\}$ with the {\em conformal} time
$t_{c}=-\frac{1}{\omega}\,e^{-\omega t}$  where the line element
\begin{equation}
ds^{2}=\frac{1}{(\omega t_c)^2}\,\left(dt_c^{2}- d{\bf x}\cdot d{\bf
x}\right)\,,
\end{equation}
is the conformal transformation of the Minkowski one \cite{rf:8}. We denote by
$A$ the Maxwell field in the chart of proper time and by $\tilde A$ the same
field in the chart of conformal time. Obviously, the field components in these
charts are related through
\begin{equation}\label{c}
A_i=\tilde A_i\,, \quad A_0=-\omega t_c \tilde A_0\,.
\end{equation}

The $SO(4,1)$ group of the pseudo-orthogonal transformations in
$M^5$ constitutes the isometry group of $M$. The basis-generators of
the $so(4,1)$ algebra are associated to ten independent Killing
vectors, ${K}_{(AB)}=-{K}_{(BA)}$, which give rise to the
basis-generators $X_{(AB)}$ of the vector representation of the
$so(4,1)$ algebra carried by the space of the vector potential, $A$.
This algebra yields the principal observables, i. e.  the energy
operator $H=\omega X_{(05)}$, the momentum components
$P^i=\omega(X_{(5i)}-X_{(0i)})$ and those of the total angular
momentum $J_i=\frac{1}{2}\varepsilon_{ijk}X_{(jk)}$
($i,j,...=1,2,3$) \cite{rf:7}. The action of these operators can be
calculated according to Eq. (\ref{XA}) using the concrete form of
the corresponding Killing vectors whose components in the chart
$\{t_c,{\bf x}\}$ read: $ K^{\mu}_{(05)}=x^{\mu}$ and
\begin{eqnarray}
K^0_{(5i)}-K^0_{(0i)}=0\,,& \quad &K^j_{(5i)}-K^j_{(0i)}=
\frac{1}{\omega}\,\delta_{ij}\,,\\
K^0_{(ij)}=0\,,& &K^k_{(ij)}=\delta_{ki}x^j-\delta_{kj}x^i\,.
\end{eqnarray}
The energy and momentum operators do not have spin parts, acting as
\begin{eqnarray}
(H\, \tilde A)_{\mu}(t_c,\bf{x})&=&-i\omega(t_c \partial_{t_c} + x^i\partial_i
+1
) {\tilde A}_{\mu}(t_c,\bf{x})\,,\label{HH}\\
(P^i \tilde A)_{\mu}(t_c,\bf{x})&=&-i\partial_i\,\tilde
A_{\mu}(t,\bf{x})\,,\label{PP}
\end{eqnarray}
while the action of the total angular momentum reads
\begin{eqnarray}
(J_i\, \tilde A)_j(t_c,\bf{x})&=&(L_i \tilde A)_j(t_c,{\bf x})
-i\varepsilon_{ijk}\tilde A_k(t_c,{\bf x})\,,\label{J}\\
(J_i\, \tilde A)_0(t_c,\bf{x})&=&(L_i \tilde A)_0(t_c,{\bf x})\,,
\end{eqnarray}
where ${\bf L}={\bf x}\times {\bf P}$ is the usual angular momentum operator.
In addition, we define the Pauli-Lubanski (or helicity) operator $W={\bf
P}\cdot {\bf J}$ whose action depends only on the spin parts,
\begin{equation}
(W \tilde A)_i(t_c,{\bf x})=\varepsilon_{ijk}\partial_j \tilde A_k(t_c,{\bf
x})\,, \quad (W \tilde A)_0(t_c,{\bf x})=0\,.
\end{equation}
This operator will define the  polarization in the canonical basis of the
$so(3)$ algebra as in special relativity.

In the chart $\{t_c,{\bf x}\}$ the conserved quantities (\ref{Ck})
can be written as
\begin{equation}\label{Ck1}
C[{K}_{(AB)}]=\frac{1}{2}\,\left\{ \delta_{ij} \left<\tilde A_i,[X_{(AB)}\tilde
A]_j\right>- \left<\tilde A_0,[X_{(AB)}\tilde A]_0\right> \right\}
\end{equation}
using the new notation
\begin{equation}\label{fg}
\left<f,g\right>=i\int d^3x\, f^*(t_c,{\bf x})
\stackrel{\leftrightarrow\,\,\,}{\partial_{t_c}} g(t_c,{\bf x})\,.
\end{equation}
The integral (\ref{fg}) defines a Hermitian form which has to play a
similar role as the relativistic scalar products of the Dirac
\cite{rf:7}, Proca \cite{rf:13} and scalar \cite{rf:11} charged fields.

The other chart, $\{t,{\bf x}\}$, is suitable for analyzing the  two
time-evolution pictures we need here \cite{rf:10}. The first one is
the natural picture (NP) which is the genuine theory as resulting
from the action (\ref{action}). The second picture, we called Schr\"
odinger picture (SP), is derived from the NP using the
transformation $A(x)\to A^S(x)=U(x)A(x)U^{-1}(x)$ produced by the
operator of time-dependent {\em dilatations} \cite{rf:10}
\begin{equation}\label{U}
U(x)=\exp\left[-\omega t(x^i \partial_i)\right]\,,
\end{equation}
which has the following convenient action
\begin{equation}\label{propU}
U(x)F({x}^i)U^{-1}(x)=F\left(e^{-\omega t}{x}^i\right)\,,\quad
U(x)G(\partial_i)U^{-1}(x)=G\left(e^{\omega t}\partial_i\right)\,,
\end{equation}
upon any analytical functions $F$ and $G$. In this new picture the conserved
operators $H_S=U(x)HU(x)^{-1}$ and $P^i_S=U(x)P^iU(x)^{-1}$ act as
\begin{eqnarray}
(H_S\, A^S)_{i}(t,\bf{x})&=&i(\partial_t -\omega ) {A}^S_{i}(t,\bf{x})\,,\label{HHS}\\
(H_S\, A^S)_{0}(t,\bf{x})&=&i\partial_t  {A}^S_{0}(t,\bf{x})\,,\\
(P^i_S A^S)_{\mu}(t,\bf{x})&=&-i e^{\omega t}
\partial_i\,A^S_{\mu}(t,\bf{x})\,,\label{PPS}
\end{eqnarray}
while the Pauli-Lubanski operator remains unchanged since it
commutes with $U$. We shall show that the SP is the suitable
framework for deriving the quantum modes of the energy basis in the
same manner as in Refs. \cite{rf:11} and \cite{rf:12}.

\section{Polarized plane wave solutions}

The specific feature of the quantum mechanics on $M$ is that the
energy operator (\ref{HH}) does not commute with the momentum
components (\ref{PP}). Therefore, there are no particular solutions
of the field equation with well-determined energy and momentum and,
consequently, we can not speak about mass-shells. This leads us to
consider different sets of plane waves solutions, depending either
on momentum or on energy and momentum direction, determining thus
two different bases, namely the momentum and energy ones.

For both these bases we consider the Coulomb gauge,
\begin{equation}
A_0=\tilde A_0=0\,,\quad \partial_i A_i=0\,,
\end{equation}
since this is {\em invariant} under the conformal transformation
which relates the Minkowski metric to that of the moving  chart
$\{t_c,{\bf x}\}$. This means that the  theory of the Maxwell field
in this chart and Coulomb gauge can be conformally-related to the
usual electrodynamic of special relativity from which we can take
over the basic results. In what follows we adopt this conjecture
assuming, in addition, that the polarization is circular.

\subsection{The momentum basis}

Thanks to this conformal invariance, the field equation
$(\partial_{t_c}^2-\Delta)A_i=0$ in the chart $\{t_{c},\bf{x}\}$ and
Coulomb gauge  has the same form as in the Minkowski case.
Therefore, the non-vanishing components of $A$ can be expanded as
\begin{eqnarray}
A_i(x)&=&A_i^{(+)}(x)+A_i^{(-)}(x)\nonumber\\
&=&\int d^3k \sum_{\lambda}\left[e_i({\bf n}_k,\lambda) f_{\bf k}(x) a({\bf
k},\lambda)+[e_i({\bf n}_k,\lambda)f_{\bf k}(x)]^* a^*({\bf
k},\lambda)\right]\,,\label{field1}
\end{eqnarray}
in terms of wave functions in momentum representation, $a({\bf k},\lambda)$,
polarization vectors, $e_i({\bf n}_k,\lambda)$, and fundamental solutions of
the d'Alambert equation,
\begin{equation}\label{fk}
f_{\bf k}(x)=\frac{1}{(2\pi)^{3/2}}\frac{1}{\sqrt{2k}}\,e^{-ikt_c+i{\bf k}\cdot
{\bf x}}\,,
\end{equation}
where ${\bf k}=k {\bf n}_k$ is the momentum vector and $k=|{\bf k}|$. The
functions  $f_{\bf k}(x)$  are assumed to be of positive frequencies while
those of negative frequencies are $f_{\bf k}(x)^*$. These solutions satisfy the
orthonormalization relations
\begin{eqnarray}
\langle  f_{\bf k},f_{{\bf k}'}\rangle=-\,\langle  f_{\bf k}^*,f_{{\bf
k}'}^*\rangle&=&\delta^3({\bf k}-{\bf k}')\,,\\
\langle f_{\bf k},f_{{\bf k}'}^*\rangle&=&0\,,
\end{eqnarray}
and the completeness condition
\begin{equation}\label{comp}
i\int d^3k\,  f^*_{\bf k}(t_c,{\bf x})
\stackrel{\leftrightarrow\,\,}{\partial_{t_c}} f_{\bf k}(t_c,{\bf
x}')=\delta^3({\bf x}-{\bf x}')\,,
\end{equation}
with respect to the Hermitian form (\ref{fg}) which plays thus the role of a
generalized scalar product as we observed before.

The polarization vectors ${\bf e}({\bf n}_k,\lambda)$ in Coulomb gauge must be
orthogonal to the momentum direction,
\begin{equation}
{\bf k}\cdot{\bf e}({\bf n}_k,\lambda)=0\,,
\end{equation}
for any polarization $\lambda=\pm 1$. We remind the reader that the
polarization can be defined in different manners independent of the
form of the scalar solutions $f_{\bf k}$. In general, the
polarization vectors have c-number components which must satisfy
\cite{rf:19}
\begin{eqnarray}
{\bf e}({\bf n}_k,\lambda)\cdot{\bf e}({\bf
n}_k,\lambda')^*&=&\delta_{\lambda\lambda'}\,,\\
 \sum_{\lambda}e_i({\bf n}_k,\lambda)\,e_j({\bf
n}_k,\lambda)^*&=&\delta_{ij}-\frac{k^i k^j}{k^2}\,.\label{tran}
\end{eqnarray}
Here we restrict ourselves to consider only the {\em circular} polarization for
which the supplementary condition ${\bf e}({\bf n}_k,\lambda)^*\land {\bf
e}({\bf n}_k,\lambda)=i \lambda\, {\bf n}_k$ is requested.

We obtained thus well-defined mode functions which represent
transverse plane waves of given momentum and helicity. Those of
positive frequencies, $w_{i({\bf k},\lambda)}=e_i({\bf n}_k,\lambda)
f_{\bf k}$, are common eigenfunctions of the complete set of
commuting operators $\{P^i,W\}$ corresponding to the eigenvalues
$\{k^i,k\lambda\}$ where $\lambda=\pm 1$. The negative frequency
plane waves, $w_{i({\bf k},\lambda)}^*$, are eigenfunctions of the
same set of operators but corresponding to the eigenvalues
$\{-k^i,-k\lambda\}$. We say that these sets of fundamental
solutions of the Maxwell equation define the {\em momentum} basis.

We specify that our definition of the positive and negative
frequency modes is similar to that of special relativity,  selecting
thus the {\em conformal} vacuum of the Maxwell theory on the dS
spacetime. Moreover, we observe that this vacuum is of the
Bunch-Davies type \cite{rf:14}. This result is not surprising as long
as the conformal vacuum of the massless scalar field conformally
coupled to the dS gravity is just the Bunch-Davies one. Moreover,
the invariance under translations requires this vacuum to be stable
\cite{rf:8}.

\subsection{The energy basis}

In the chart $\{t,\bf{x}\}$ the potential in Coulomb gauge has the same
components, $A_i$, which satisfy the field equation of the NP,
\begin{equation}
\partial_t^2A_i+\omega \partial_t A_i-e^{-2\omega t}\Delta A_i=0\,.
\end{equation}
The operator (\ref{U}) transforms this equation into the field equation of the
SP,
\begin{equation}\label{FE2}
\left[\left( \partial_t+\omega x^i\partial_i \right)^2+\omega
(\partial_t+\omega x^i\partial_i)-\Delta \right]A^S_i=0\,,
\end{equation}
which does not depend explicitly on time. We show that this equation has
particular solutions representing plane waves of given energy, momentum
direction and polarization.

Let us start assuming that in the SP the  potential can be expanded as
\begin{eqnarray}
A^S_i(x)&=&A_i^{S\,(+)}(x)+A_i^{S\,(-)}(x)\nonumber\\
& =&\int_{0}^{\infty} dE \int d^3q\, \left[\hat A_i^{S\,(+)}(E,{\bf
q})e^{-i(Et-{\bf q}\cdot{\bf x})}\right.\nonumber\\
&&\hspace*{24mm}\left.+\, \hat A_i^{S\,(-)}(E,{\bf q})e^{i(Et-{\bf q}\cdot{\bf
x})}\right]e^{\omega t}\label{KGS}
\end{eqnarray}
where $E\ge 0$ is the energy defined as the eigenvalue of the operator $H_S$
which acts as in Eq. (\ref{HHS}). Whenever the fields $\hat{A}^{S\,(\pm)}_i$
behave as tempered distributions on the domain ${\Bbb R}_q^3$, the Green
theorem may be used for replacing the new momentum operators  $-i\partial_i$ of
the SP by ${q}^i$ and the coordinates $x^i$ by $i {\partial}_{q_i}$ obtaining
thus the field equation of the SP in momentum representation,
\begin{eqnarray}
&&\left\{\left[\pm i E+\omega \left({q}^i{
\partial}_{q_i}+2\right)\right]^2\right.\nonumber\\
&&~~~~~~~~\left. -\, \omega \left[\pm i E+\omega \left({q}^i{
\partial}_{q_i}+2\right)\right]+{\bf q}^2\right\}
\hat A_i^{S\,(\pm)}(E,{\bf q})=0\,.\label{KG4}
\end{eqnarray}

The energy $E$ is a conserved quantity but the momentum ${\bf q}$
does not have this property since the operators $-i\partial_i$ of
the SP are no longer produced by Killing vectors. More specific,
only the scalar momentum $q=|{\bf q}|$ is not conserved while the
momentum direction is conserved since the operator $-i\nabla$  is
parallel with the conserved momentum ${\bf P}^S$ given by Eq.
(\ref{PPS}). For this reason we denote ${\bf q}=q\, {\bf n}$
observing that the differential operator of Eq. (\ref{KG4}) is of
radial type and reads ${q}^i{\partial}_{q_i}=q\,\partial_q$.
Consequently, this operator acts only on the functions depending on
$q$ while the functions which depend on the momentum direction ${\bf
n}$ behave as constants. Therefore, we have to look for solutions of
the form
\begin{equation}
\hat A_i^{S\,(+)}(E,{\bf q})=[\hat A_i^{S\,(-)}(E,{\bf
q})]^*=h_S(E,{q})\,e_i({\bf n},\lambda) \,a(E,{\bf n},\lambda)\,,
\end{equation}
where the function $h_S$ satisfies an equation derived from Eq.
(\ref{KG4}) that can be written simply as
\begin{equation}
\left[\frac{d^2}{ds^2}+\frac{2i\epsilon+4}{s}\frac{d}{ds}
+\frac{3i\epsilon+2-\epsilon^2}{s^2}+1\right] h_S(\epsilon, s)=0
\end{equation}
using the new variable $s=q/\omega$ and the notation
$\epsilon=E/\omega$. This equation has solutions of the form
$s^{-i\epsilon -2}\,e^{\pm i s}$. Observing that the positive
frequency solutions (\ref{fk}) are  progressive plane waves, we
choose $h_S(\epsilon,s)={\rm const}\,\, s^{-i\epsilon -2}\,e^{i s}$
in order to obtain positive frequency energy eigenfunctions of the
same type as in Eq. (\ref{fdef}). This choice guarantees the
uniqueness of the vacuum for both the bases we consider here.

Collecting now all the above results we can rewrite  the field
(\ref{KGS})  as
\begin{eqnarray}
A_i^S(x)&=&\int_0^{\infty}\,dE\int_{S^2}\, d\Omega_n\,\sum_{\lambda}\,
\left\{e_i({\bf
n},\lambda) f^S_{E,{\bf n}}(x) a(E,{\bf n},\lambda)\right.\nonumber\\
&& \hspace*{28mm}+ \, \left.[e_i({\bf n},\lambda) f^S_{E,{\bf n}}(x)]^*
a^*(E,{\bf n},\lambda)\right\}\,,
\end{eqnarray}
bearing in mind that the second integration covers the sphere
$S^2\subset {\Bbb R}^3_p$. The functions $f^S_{E,{\bf n}}$ of
positive frequencies, energy $E$ and momentum direction ${\bf n}$
have the integral representation
\begin{equation}\label{fps}
f^S_{E,{\bf n}}(x)=
 Ne^{-iEt+\omega t}\int_{0}^{\infty} ds\,
 e^{i s+ i \omega s {\bf n}\cdot{\bf x}-i\epsilon \ln s}\,,
\end{equation}
where $N$ is a normalization constant.

The physical meaning of this result may be pointed out turning back
to the NP. In this picture the  potential
\begin{eqnarray}
A_i(x)&=&\int_0^{\infty}\,dE\int_{S^2}\, d\Omega_n\,\sum_{\lambda}\,
\left\{e_i({\bf
n},\lambda) f_{E,{\bf n}}(x) a(E,{\bf n},\lambda)\right.\nonumber\\
&& \hspace*{28mm} + \, \left.[e_i({\bf n},\lambda) f_{E,{\bf n}}(x)]^*
a^*(E,{\bf n},\lambda)\right\}\,,\label{field2}
\end{eqnarray}
is expressed in terms of the scalar functions of the NP that read
\begin{equation}\label{fpsNP}
f_{E,{\bf n}}({x})=U^{-1}(x)f^S_{E,{\bf n}}({x})U(x)=
 Ne^{-iEt+\omega t}\int_{0}^{\infty} ds\,
  e^{i s + i \omega s {\bf n}\cdot{\bf x}_t-i\epsilon \ln
 s}\,,
\end{equation}
where ${\bf x}_t=e^{\omega t}{\bf x}$. Finally, changing the
integration variable, $e^{\omega t}s\to s$, we obtain the definitive
result
\begin{equation}\label{fdef}
f_{E,{\bf n}}(t_c,{\bf x})=
 N \int_{0}^{\infty} ds\,
  e^{i \omega s ({\bf n}\cdot{\bf x}-t_c)-i\epsilon \ln
 s}=N \,\Gamma(1-i\epsilon)(it_c-i\,{\bf n}\cdot{\bf x}+0)^{i\epsilon-1}\,,
\end{equation}
but in the chart of the conformal time $t_c$. Since these functions
are of positive frequencies we define the negative frequency ones as
being $f_{E,{\bf n}}(t_c,{\bf x})^*$. Notice that all these
functions are regular inside the light-cone, including the
null-geodesics where $\omega({\bf n}\cdot{\bf x}-t_c)=1$.

Using now the Hermitian form (\ref{fg}) we can prove (as in Appendix) that the
normalization constant
\begin{equation}\label{norm}
N=\frac{1}{2} \sqrt{\frac{\omega}{\pi}}\frac{1}{(2\pi)^{3/2}}
\end{equation}
(defined up to a phase factor) assures the desired orthonormalization
relations,
\begin{eqnarray}
 \langle f_{E,{\bf n}},f_{E',{\bf
n}^{\,\prime}}\rangle=- \langle f^*_{E,{\bf n}},f^*_{E',{\bf
n}^{\,\prime}}\rangle&=& \delta(E-E')\,\delta^2 ({\bf n}-{\bf n}^{\,\prime})\,,
\label{orto1}\\
\langle f_{E,{\bf n}},f^*_{E',{\bf n}^{\,\prime}}\rangle&=&0\,, \label{orto2}
\end{eqnarray}
and the completeness condition
\begin{equation}\label{comp1}
i\int_0^{\infty}dE\int_{S^2} d\Omega_n \left\{ [f_{E,{\bf n}}(t_c,{\bf x})]^*
\stackrel{\leftrightarrow}{\,\,\partial_{t_c}} f_{E,{\bf n}}(t_c,{\bf x}')
\right\} =\delta^3 ({\bf x}-{\bf x}^{\,\prime})\,.
\end{equation}
As expected, Eqs. (\ref{orto1}) and (\ref{orto2}) show that the functions of
positive and negative frequencies are orthogonal to each other.

The plane waves $w_{i(E,{\bf n},\lambda)}=e_i({\bf n},\lambda) f_{E,\bf n}$
depend on the energy $E$, that is the eigenvalue of $H$, and on the direction
${\bf n}$ and the polarization $\lambda$ which are no longer eigenvalues of
differential operators. For this reason,  the complete set of commuting
operators determining the energy basis can be defined only at the level of
quantum field theory. Nevertheless, we observe that the plane waves of negative
frequencies, $w_{i(E,{\bf n},\lambda)}^*$ are eigenfunctions of the operator
$H$ corresponding to the eigenvalue $-E$. Therefore, we can say that these sets
of plane waves constitute the complete system of fundamental solutions of the
Maxwell equation defining the {\em energy} basis of the NP.

Working simultaneously with two bases (but in the same chart) we need to know
the transition coefficients which can be calculated straightforwardly as
\begin{eqnarray}
&&\langle f_{\bf k},f_{E,{\bf n}}\rangle=-\langle f_{\bf k}^*, f^*_{E,{\bf
n}}\rangle^*=\frac{k^{-\frac{3}{2}}}{\sqrt{2\pi\omega}}\,\delta^2({\bf n}-{\bf
n}_k)\,e^{-i\frac{E}{\omega}\ln \frac{k}{\omega}}\,,\label{trcof}\\
&&\langle f_{\bf k},f_{E,{\bf n}}^*\rangle=\langle f_{\bf k}^*, f_{E,{\bf
n}}\rangle=0
\end{eqnarray}
where ${\bf n}_k={\bf k}/k$. With their help we deduce the transformations
\begin{eqnarray}
a({\bf k},\lambda)&=&\int_0^{\infty}dE\int_{S^2}d\Omega_n \langle f_{\bf
k},f_{E,{\bf
n}}\rangle a(E,{\bf n},\lambda)\nonumber\\
&=&\frac{k^{-3/2}}{\sqrt{2\pi\omega}}\int_0^{\infty}dE\,
e^{-i\frac{E}{\omega}\ln \frac{k}{\omega}}\,a(E,{\bf n}_k,\lambda)\,,\label{Iaa1}\\
a(E,{\bf n},\lambda)&=&\int d^3 k \,\langle f_{E,{\bf n}},f_{\bf k}\rangle
a({\bf
k},\lambda)\nonumber\\
&=&\frac{1}{\sqrt{2\pi\omega}}\int_0^{\infty}dk\,\sqrt{k}\,\,
e^{\,i\frac{E}{\omega}\ln \frac{k}{\omega}}\,a(k\,{\bf
n},\lambda)\,,\label{Iaa2}
\end{eqnarray}
which are similar to those found for the scalar field \cite{rf:11}.
These relations define the unitary transformation (in the
generalized sense) between the momentum and energy bases.

Finally, we must stress that the separation of the positive and
negative frequency modes of the momentum and respectively energy
bases we adopted here leads to the {\em same} vacuum  which is just
the conformal vacuum of the Bunch-Davies type.  For this reason the
transformations (\ref{Iaa1}) and (\ref{Iaa2}) do not mix the
particle and antiparticle subspaces between themselves and,
therefeore, can not be interpreted as Bogoliubov transformations
giving rise to the Unruh \cite{rf:20,rf:21} or Gibbons-Hawking \cite{rf:22}
effects.

\section{Quantization in Coulomb gauge}

The conformal invariance of the whole theory in Coulomb gauge
enables us to perform the second quantization in canonical manner as
in special relativity. We assume that the wave functions $a$ of the
fields (\ref{field1}) and (\ref{field2}) become field operators
(with $a^{*}\to a^{\dagger}$) \cite{rf:19} which fulfill the standard
commutation relations in the momentum basis from which the
non-vanishing ones are
\begin{equation}\label{com1}
[a({\bf k},\lambda), a^{\dagger}({\bf k}^{\,\prime},\lambda ')]
=\delta_{\lambda\lambda '} \delta^3 ({\bf k}-{\bf k}^{\,\prime})\,.
\end{equation}
Then, from Eq. (\ref{Iaa1}) it results that the field operators of the energy
basis satisfy
\begin{equation}\label{com2}
[a(E,{\bf n},\lambda), a^{\dagger}(E',{\bf n}^{\,\prime},\lambda
')]=\delta_{\lambda\lambda '} \delta(E-E') \delta^2 ({\bf n}-{\bf
n}^{\,\prime})\,,
\end{equation}
and
\begin{equation}
[a({\bf p},\lambda), a^{\dagger}(E,{\bf n},\lambda)]=\langle f_{\bf
p},f_{E,{\bf n}}\rangle\,,
\end{equation}
while other commutators are vanishing.

The  Hermitian field $A=A^{\dagger}$ is now correctly quantized
according to the {\em canonical} rule
\begin{equation}
[ A_i(t_c,{\bf x}),\pi^j(t_c,{\bf x}')]=[ A_i(t_c,{\bf
x}),\partial_{t_c}A_j(t_c,{\bf x}')]=i\,\delta^{tr}_{ij}({\bf x}-{\bf x}')\,,
\end{equation}
where
\begin{equation}
\pi^j=\sqrt{g}\,\frac{\delta {\cal L}}{\delta
(\partial_{t_c}A_j)}=\partial_{t_c}A_j
\end{equation}
is the momentum density in Coulomb gauge ($A_0=0$) and
\begin{equation}
\delta^{tr}_{ij}({\bf x})=\frac{1}{(2\pi)^3}\int d^3q
\left(\delta_{ij}-\frac{q^iq^j}{q^2}\right)e^{i {\bf q}\cdot {\bf x}}
\end{equation}
is the well-known transverse $\delta$-function \cite{rf:19} arising
from Eq. (\ref{tran}).

We remark the advantage of the Coulomb gauge which helps us to take
over the well-known results of special relativity including the
commutation relations \cite{rf:19}. However, in general, the Coulomb
gauge is no mandatory, other methods of special relativity being
able to be adopted for quantizing the Maxwell field in various
gauges on dS or even FRW manifolds \cite{rf:23}.

What is new in our approach is the presence of the momentum and
energy bases which will generate two corresponding different bases
of the Fock space. The arguments we presented above indicate that
the vacuum state $|0\rangle$ of the Fock space is unique and
well-defined. Therefore, the field operators have the usual action,
\begin{equation}
a({\bf k},\lambda)\,|0\rangle=0\,,\quad \langle 0|\,a^{\dagger}({\bf
k},\lambda)=0\,,
\end{equation}
and similarly for the operators of the energy basis. The sectors with a given
number of particles have to be constructed using the standard methods,
obtaining thus the generalized bases of momentum or energy.

Important pieces of the quantum theory are the Green functions related to the
partial commutator functions (of positive or negative frequencies) defined as
\begin{equation}
D_{ij}^{(\pm)}(x-x')= i[A_i^{(\pm)}(x),A_j^{(\pm)\,\dagger}(x')]
\end{equation}
and the total one, $D_{ij}=D^{(+)}_{ij}+D^{(-)}_{ij}$. These function are
solutions of the field equation with vanishing divergences in both the sets of
variables and obey $[D_{ij}^{(\pm)}]^*=D_{ij}^{(\mp)}$  such that $D_{ij}$
results to be a real function. Thus it is enough to focus only on the functions
of positive frequencies,
\begin{eqnarray}
D^{(+)}_{ij}(x-x')&=&i\int d^3 k \, f_{\bf k}(x)f_{\bf
k}(x')^*\left(\delta_{ij}-\frac{k^i k^j}{k^2}\right)\nonumber\\
&=&i\int_0^{\infty} dE \int_{S^2} d\Omega_n \, f_{E,\bf
n}(x)f_{E,\bf n}(x')^*\left(\delta_{ij}-n^i n^j\right)\,,
\end{eqnarray}
resulted from  Eqs. (\ref{field1}), (\ref{field2}) and (\ref{tran}). Both these
versions lead to the final expression
\begin{equation}
D_{ij}^{(+)}(x-x')=\frac{i}{(2\pi)^3}\,  \int \frac{d^3k}{2 k} \,
\left(\delta_{ij}-\frac{k^i k^j}{k^2}\right)e^{i{\bf k}\cdot({\bf
x}-{\bf x}')-i k(t_c-t_c')}
\end{equation}
from which we deduce what happens at equal time,
\begin{equation}\label{Dtt}
\left.\partial_{t_c}D_{ij}^{(+)}(t_c-t_c',{\bf x}-{\bf
x}')\right|_{t_c'=t_c}=\frac{1}{2}\,\delta^{tr}_{ij}({\bf x}-{\bf
x}')\,.
\end{equation}

Using such functions we can construct different {\em transverse}
Green functions, $G_{ij}(x)=G_{ji}(x)$, which obey
\begin{equation}\label{KGG}
\left( \partial_{t_c}^2-\Delta_x
\right)G_{ij}(x-x')=\delta(t_c-t_c')\delta^{tr}_{ij}({\bf x}-{\bf
x}')
\end{equation}
and $\partial_{i}G_{\cdot j}^{i \cdot}(x)=0$. Of a special interest
are the retarded, $D^R_{ij}(x)=\theta(t_c)D_{ij}(x)$, and advanced,
$D^A_{ij}(x)=-\theta(-t_c)D_{ij}(x)$, transverse Green functions.
The transverse Feynman propagator,
\begin{eqnarray}
&&D^F_{ij}(x-x')= i\langle 0|T[A_i(x) A_j(x')]\,|0\rangle\nonumber\\
&&= \theta (t_c-t_c')
D_{ij}^{(+)}(x-x')-\theta(t_c'-t_c)D_{ij}^{(-)}(x- x')\,,
\end{eqnarray}
is defined as a  causal Green function. It is not difficult to verify that all
these functions satisfy Eq. (\ref{KGG}) if one uses the identity
$\partial_t^2[\theta(t)f(t)]=\delta(t)\partial_t f(t)$ and Eq. (\ref{Dtt}).

The conclusion is that the Green functions of the chart
$\{t_{c},\bf{x}\}$  have the same forms and properties as those of
the Maxwell theory in the Minkowski spacetime, including the
representation in the complex $k_0$-plane. The difference is that
here the particular value $k_0=k$ can not be interpreted as the
photon energy since there are no mass-shells. However, these
similarities are merely formal while the physical meaning is quite
different because of the special definition of the conformal time.

\section{One-particle operators}

The one-particle operators corresponding to the conserved quantities
(\ref{Ck1}) can be calculated in the Coulomb gauge as
\begin{equation}\label{opo}
{\cal X}=\frac{1}{2}\,\delta_{ij}:\langle A_i, (X A)_j\rangle:
\end{equation}
respecting the normal ordering of the operator products \cite{rf:19}. The obvious
algebraic properties
\begin{equation}\label{algXX}
[{\cal X}, A_i(x))]=-(X A)_i(x)\,, \quad [{\cal X}, {\cal
Y}\,]=\frac{1}{2}\,\delta_{ij}:\langle A_i, ([X,Y]A)_j\,\rangle:
\end{equation}
are due to the canonical quantization adopted here. However, there are many
other conserved operators which do not have corresponding differential
operators at the level of the relativistic quantum mechanics.  The simplest
example is the operator of the number of particles,
\begin{equation}
{\cal N}=\int d^3k \sum_{\lambda}   a^{\dagger}({\bf k},\lambda) a({\bf
k},\lambda)=\int_0^{\infty}dE \int_{S^2}  d\Omega_n \sum_{\lambda}
a^{\dagger}(E,{\bf n},\lambda) a(E,{\bf n},\lambda)\,,
\end{equation}

The principal conserved one-particle operators are the components of momentum
operator,
\begin{equation}
{\cal P}^l=\frac{1}{2}\,\delta_{ij}:\langle A_i,  (P^l A)_j\rangle:=\int d^3k\,
k^l \sum_{\lambda}  a^{\dagger}({\bf k},\lambda) a({\bf k},\lambda)\,,
\end{equation}
and the Pauli-Lubanski operator,
\begin{equation}
{\cal W}=\frac{1}{2}\,\delta_{ij}:\langle A_i, (W A)_j\rangle:\,=\int d^3k\,k
 \sum_{\lambda}\lambda\,  a^{\dagger}({\bf k},\lambda) a({\bf k},\lambda)\,,
\end{equation}
which are diagonal in the momentum basis as well as the energy operator,
\begin{equation}
{\cal H}=\frac{1}{2}\,\delta_{ij}:\langle A_i, (H A)_j\rangle:=\int_0^{\infty}
dE\, E \int_{S^2} d\Omega_n \sum_{\lambda} a^{\dagger}(E,{\bf n},\lambda)
a(E,{\bf n},\lambda)\,,
\end{equation}
expanded in the energy basis.

More interesting are the operators $\tilde{\cal P}^i$ of the momentum direction
since they do not come from differential operators and, therefore, must be
defined directly as
\begin{equation}
\tilde{\cal P}^i=\int_0^{\infty} dE \int_{S^2} d\Omega_n n^i\,\sum_{\lambda}\,
a^{\dagger}(E,{\bf n},\lambda) a(E,{\bf n},\lambda)\,.
\end{equation}
Similarly we define the new {\em normalized} Pauli-Lubanski operator
\begin{eqnarray}
\tilde {\cal W}&=&\frac{1}{2}\,\delta_{ij}:\langle A_i, (W A)_j\rangle:\,=\int
d^3k
 \sum_{\lambda}\lambda\,  a^{\dagger}({\bf k},\lambda) a({\bf k},\lambda)\nonumber\\
&=&\int_0^{\infty} dE\, \int_{S^2} d\Omega_n \sum_{\lambda} \lambda
\,a^{\dagger}(E,{\bf n},\lambda) a(E,{\bf n},\lambda)\,,
\end{eqnarray}
which is diagonal in both our bases.

The above operators which satisfy simple commutation relations,
\begin{eqnarray}\label{comHP}
&&[{\cal H}, {\cal P}^i]=i\omega {\cal P}^i\,,\quad [{\cal H}, {\cal
W}]=i\omega {\cal W}\,,\quad [{\cal H},\tilde{\cal P}^i]=
[{\cal H}, \tilde{\cal W}]=0\,,\\
 && \hspace*{12mm}[{\cal W}, {\cal P}^i]=[{\cal W},\tilde {\cal P}^i]=
 [\tilde{\cal W}, {\cal P}^i]=[\tilde{\cal W},\tilde {\cal P}^i]=0\,,
\end{eqnarray}
determine the momentum and energy bases as  common eigenvectors of
the sets of commuting operators  $\{{\cal P}^i,  {\cal W}\}$ and
respectively $\{{\cal H},\tilde {\cal P}^i,\tilde {\cal W}\}$.

It is worth pointing out that the transition coefficients (\ref{trcof}) can be
used for finding closed expressions for conserved one-particle operators in
bases in which these operators are not diagonal. For example, we can calculate
the energy operator in the momentum basis either starting with the identity
\begin{equation}
(Hf_{\bf k})(x)=-i\omega \left(k^i\partial_{k_i}+{\frac{3}{2}}\right)f_{\bf
k}(x)
\end{equation}
or by using directly Eq. (\ref{Iaa2}). The final result,
\begin{equation}\label{Hpp}
{\cal H}=\frac{i\omega}{2}\int d^3k\, k^i  \sum_{\lambda}\,
a^{\dagger}({\bf
k},\lambda)\stackrel{\leftrightarrow}{\partial}_{k_i} a({\bf
k},\lambda)\,,
\end{equation}
is similar to that obtained for the scalar \cite{rf:11} and  Dirac
\cite{rf:7} fields on $M$.

The components of total angular momentum are not diagonal in the above
considered bases but they can be easily represented in both of these bases.
According to Eqs. (\ref{XA}) and (\ref{J}), we find the following expansion in
the momentum basis:
\begin{eqnarray}
{\cal J}_l&=&-\frac{i}{2}\,\varepsilon_{lij}\int d^3k\,\left[ k^i
\sum_{\lambda}\, a^{\dagger}({\bf
k},\lambda)\stackrel{\leftrightarrow}{\partial}_{k_j} a({\bf
k},\lambda)\right.\nonumber\\
&&+\left.\sum_{\lambda \lambda'}\, \vartheta^{ij}_{\lambda
\lambda'}({\bf k})\,a^{\dagger}({\bf k},\lambda) a({\bf
k},\lambda')\right]\,,
\end{eqnarray}
where
\begin{equation}
\vartheta^{ij}_{\lambda \lambda'}({\bf k})=2 e_i({\bf
k},\lambda)^*e_j({\bf k},\lambda')+ k^i\sum_l e_l({\bf
k},\lambda)^*\stackrel{\leftrightarrow}{\partial}_{k_j} e_l({\bf
k},\lambda')\,,
\end{equation}
and we recover the identity ${\cal W}=\sum_i{\cal P}_i{\cal J}_i$.
Similar formulas can be written in the energy basis.

\section{Concluding remarks}

We presented the complete quantum theory of the Maxwell field
minimally coupled to the gravity of the dS backgrounds, working in
the moving charts of this manifold. The main points of our approach
are the method of constructing conserved operators, the new Schr\"
odinger time-evolution picture and the choice of the Coulomb gauge
in which the whole theory can be conformally-related to the flat
case. Under such circumstances, the principal results in the chart
of conformal time are given by similar formulas as in special
relativity. However, the physical meaning is different since in the
flat limit the conformal time does not tend to the Minkowski one.

We derived the mode functions as solutions of the field equation
which are common eigenfunctions of some complete sets of commuting
operators that commute with the equation operator too. All these
operators are globally defined on the whole dS manifold, having
global algebraic properties, but making sense in a given chart only
in the observer's light-cone where this can perform physical
measurements. Moreover, there exists a relativistic scalar product,
globally defined, with respect to which the subspaces of the
positive and respectively negative frequency modes are orthogonal to
each other. Thus we can say that these quantum modes are prepared by
a {\em global} apparatus which is no longer a simple local detector
\cite{rf:8}.

The plane waves of the momentum and energy bases are determined as
common eigenvectors of the sets $\{{\cal P}^i,  {\cal W}\}$ and
respectively $\{{\cal H},\tilde {\cal P}^i,\tilde {\cal W}\}$. Other
spherical modes defined as eigenvectors of the set $\{{\cal H},
{\cal J}^2, {\cal J}_3 \}$  were derived in the dS static chart with
spherical coordinates \cite{rf:5}. The last two  sets of commuting
operators include the same energy operator which helps one to
separate the positive and negative frequencies. This means that the
vacuum state is the same in all these cases and, therefore, these
three types of modes can be transformed among themselves without to
mix the particle and antiparticle subspaces. We showed here how the
first two types of modes can be transformed to each other but the
transformations among the third type of modes and the first two ones
remain to be found. This could be done in two steps namely, by
writing first the spherical modes in the moving charts with
spherical coordinates and exploiting then the well-known relation
between the plane and spherical waves for relating the spherical
modes to those of the energy basis.

Finally, we remark that the canonical quantization of all the fields
we worked out so far on dS manifolds leads to quantum fields which
can be manipulated similarly as those of special relativity. This
indicates that our approach could be the starting point for building
a simple version of perturbation theory of the interacting quantum
fields on the dS expanding universe.

\appendix

\subsection*{Appendix:  Normalization integrals}

In spherical coordinates of the momentum space, ${\bf n}\sim
(\theta_n,\phi_n)$, and the notation ${\bf q}=\omega s{\bf n}$, we have $d^3q=
q^2dq\, d\Omega_n=\omega^3\, s^2ds\, d\Omega_n$ with
$d\Omega_n=d(\cos\theta_n)d\phi_n$. Moreover, we can write
\begin{equation}\label{del}
\delta^3({\bf q}-{\bf q}^{\,\prime})=\frac{1}{q^2}\,\delta(q-q')\delta^2({\bf
n}-{\bf n}') =\frac{1}{\omega^3 s^2}\,\delta(s-s')\delta^2({\bf n}-{\bf n}')\,,
\end{equation}
where we denoted $\delta^2({\bf n}-{\bf n}') =\delta(\cos \theta_n-\cos
\theta_n')\delta(\phi_n-\phi_n')\,.$

The normalization integrals can be calculated in NP starting with the Hermitian
form  (\ref{fg}). According to Eqs. (\ref{fdef}) and (\ref{del}), this yields
\begin{equation}
\langle f_{E,{\bf n}},f_{E',{\bf n}^{\,\prime}}\rangle=\frac{2 N^2
(2\pi)^3}{\omega^2}\, \delta^2({\bf n}-{\bf n}')\int_0^{\infty}
\frac{ds}{s}\,e^{i(\epsilon-\epsilon')\ln s}\,.
\end{equation}
Finally, using the identity
\begin{equation}
\frac{1}{2\pi\omega}\int_0^{\infty}\frac{ds}{s}\, e^{i(\epsilon-\epsilon')\ln
s} =\delta(E-E')\,,
\end{equation}
we find the value of the normalization constant (\ref{norm}).

Finally we note that the representation
\begin{equation}
\int_{0}^{\infty}dE e^{i\epsilon (\ln s -\ln s')}=2\pi \omega \,s\,
\delta(s-s')
\end{equation}
holds as long as all the functions we consider here depend only on positive
energies, $E\ge 0$.

\end{document}